\documentclass[a4paper,magyar,english]{article}
\usepackage{mathpazo}
\usepackage[T1]{fontenc}
\usepackage[latin2,latin9]{inputenc}
\usepackage{graphicx}
\usepackage{amssymb}

\makeatletter

\providecommand{\tabularnewline}{\\}

\newcounter {point}[section]
\renewcommand\theparagraph    {\thesection.%
\@arabic\c@paragraph}

\usepackage{babel}
\makeatother

\begin{document}

\title{The Einstein--Podolsky--Rosen Argument and the Bell Inequalities%
\thanks{To be published in \emph{The Internet Encyclopedia of Philosophy}%
}}

\author{László E. Szabó\\
\emph{\normalsize Department of Logic, Institute of Philosophy}\\
\emph{\normalsize Eötvös University, Budapest}\\
\emph{\normalsize http://phil.elte.hu/leszabo}}

\date{~}

\maketitle
In 1935 Einstein, Podolsky, and Rosen (EPR) published an important
paper \cite{key-6} in which they claimed that the whole formalism
of quantum mechanics together with what they called {}``Reality Criterion''
imply that quantum mechanics cannot be complete. That is, there must
exist some elements of reality that are not described by quantum mechanics.
There must be, they concluded, a more complete description of physical
reality behind quantum mechanics. There must be a state, a hidden
variable, characterizing the state of affairs in the world in more
details than the quantum mechanical state, something that also reflects
the missing elements of reality. 

Under some further but quite plausible assumptions, this conclusion
implies that in some spin-correlation experiments the measured quantum
mechanical probabilities should satisfy particular inequalities (Bell-type
inequalities). The paradox consists in the fact that quantum probabilities
do not satisfy these inequalities. And this paradoxical fact has been
confirmed by several laboratory experiments in the last three decades.
The problem is still open and hotly debated among both physicists
and philosophers. It has motivated a wide range of research from the
most fundamental quantum mechanical experiments through foundations
of probability theory to the theory of stochastic causality as well
as the metaphysics of free will.

\tableofcontents{}

\section{The Einstein--Podolsky--Rosen argument}

\subsection{The description of the EPR experiment\label{sub:The-description-of}}

Instead of the thought experiment described in the original EPR paper
\cite{key-6} we will formulate the problem for a more realistic spin-correlation
experiment suggested by Aharonov and Bohm \cite{key-7} in 1957.

\begin{figure}[htbp]
\begin{centering}
\includegraphics[scale=0.5]{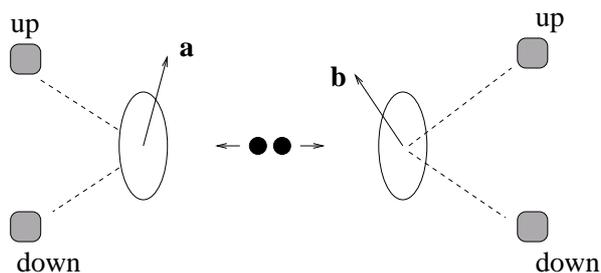}
\par\end{centering}

\caption{\emph{The Bohm--Aharonov spin-correlation experiment\label{fig_Bohm_EPR}}}

\end{figure}

Consider a source emitting two spin-$\frac{1}{2}$ particles (Fig.~\ref{fig_Bohm_EPR}).
The (spin) state space of the emitted two-particle system is $H^{2}\otimes H^{2}$,
where $H^{2}$ is a 2-dimensional Hilbert space (for a brief introduction
to quantum mechanics, see \cite{key-55}, Chapter~1). Let the quantum
state of the system be the so called \emph{singlet} state: $\hat{W}=P_{\Psi_{s}}$,
where $\Psi_{s}=\frac{1}{\sqrt{2}}\left(\psi_{+\mathbf{v}}\otimes\psi_{-\mathbf{v}}-\psi_{-\mathbf{v}}\otimes\psi_{+\mathbf{v}}\right)$.
$\psi_{+\mathbf{v}}$ and $\psi_{-\mathbf{v}}$ denote the \emph{up}
and \emph{down} eigenvectors of the spin-component operator along
an arbitrary direction $\mathbf{v}$. In the two wings, we measure
the spin-components along directions $\mathbf{a}$ and $\mathbf{b}$,
which we set up by turning the Stern--Gerlach magnets into the corresponding
positions. Let us restrict our considerations for the \emph{spin-up}
events, and introduce the following notations:\begin{eqnarray*}
A & = & \textrm{The}<\textrm{spin of the left particle is }up>\textrm{detector fires}\\
B & = & \textrm{The}<\textrm{spin of the right particle is }up>\textrm{detector fires}\\
a & = & \textrm{The left Stern--Gerlach magnet is turned into position}\mbox{ }\mathbf{a}\\
b & = & \textrm{The right Stern--Gerlach magnet is turned into position}\mbox{ }\mathbf{b}\end{eqnarray*}
In the quantum mechanical description of the experiment, events $A$
and $B$ are represented by the following subspaces of $H^{2}\otimes H^{2}$:\begin{eqnarray*}
A & = & \textrm{span}\left\{ \psi_{+\mathbf{a}}\otimes\psi_{+\mathbf{a}},\,\psi_{+\mathbf{a}}\otimes\psi_{-\mathbf{a}}\right\} \\
B & = & \textrm{span}\left\{ \psi_{+\mathbf{b}}\otimes\psi_{+\mathbf{b}},\,\psi_{+\mathbf{b}}\otimes\psi_{-\mathbf{b}}\right\} \end{eqnarray*}
 (The same capital letter $A,B,$ etc., is used for the event, for
the corresponding subspace, and for the corresponding projector, but
the context is always clear.) Quantum mechanics provides the following
probabilistic predictions: \begin{eqnarray}
p(A|a)=tr\left(P_{\Psi_{s}}A\right)=p(B|b)=tr\left(P_{\Psi_{s}}B\right) & = & \frac{1}{2}\label{eq_EPR_prob_fel}\\
p(A\wedge B|a\wedge b)=tr\left(P_{\Psi_{s}}AB\right) & = & \frac{1}{2}\sin^{2}\frac{\sphericalangle(\mathbf{a,b})}{2}\label{eq_EPR_prob_sin}\end{eqnarray}
where $\sphericalangle(\mathbf{a,b})$ denotes the angle between directions
$\mathbf{a}$ and $\mathbf{b}$. Inasmuch as we are going to deal
with sophisticated interpretational issues, the following must be
explicitly stated:

\medskip{}

\begin{center}
\begin{minipage}[c][1\totalheight]{0.9\columnwidth}%

\paragraph*{Assumption~1}

\begin{equation}
p\left(X|x\right)=tr\left(\hat{W}X\right)\label{eq:traceegyenlo}\end{equation}
That is to say, whenever we compare quantum mechanics with empirical
facts, {}``quantum probability'' $tr\left(\hat{W}X\right)$ is identified
with the conditional probability of the outcome event $X$ given that
the corresponding measurement $x$ is performed.%
\end{minipage}%

\par\end{center}

\medskip{}

\noindent This assumption is used in (\ref{eq_EPR_prob_fel})--(\ref{eq_EPR_prob_sin}). 

The two measurements happen approximately at the same time and at
two places far distant from each other. It is a generally accepted
principle in contemporary physics that there is no super-luminal propagation
of causal effects. According to this principle we have the following
assumption:

\medskip{}

\begin{center}
\begin{minipage}[c][1\totalheight]{0.9\columnwidth}%

\paragraph*{Assumption~2}

The events in the left wing (the setup of the Stern--Gerlach magnet
and the firing of the detector, etc.) cannot have causal effect on
the events in the right wing, and \emph{vice versa.}%
\end{minipage}%
 
\par\end{center}

\medskip{}

One must recognize that, in spite of this causal separation, (\ref{eq_EPR_prob_sin})
generally means that there are correlations between the outcomes of
the measurements performed in the left and in the right wings. In
particular, if $\sphericalangle(\mathbf{a,b})=0$, the correlation
is maximal: the outcome of the left measurement {}``determines'',
with probability $1$, the outcome of the right measurement. That
is, if we observe `spin-up' in the left wing then we know in advance
that the result must be `spin-down' in the right wing, and \emph{vice
versa}. The actual correlations depend on the particular measurement
setups. The very \emph{possibility} of perfect correlation is, however,
of paramount importance:

\medskip{}

\begin{center}
\begin{minipage}[c][1\totalheight]{0.9\columnwidth}%

\paragraph*{Assumption~3}

For any direction $\mathbf{b}$ in the right wing one \emph{can} chose
a direction $\mathbf{a}$ in the left wing---and \emph{vice versa}---such
that the outcome events are perfectly correlated.%
\end{minipage}%
 
\par\end{center}

\medskip{}

\subsection{The Reality Criterion}

From this fact, that the measurement outcome in the left wing {}``determines''
the outcome in the right wing, in conjunction with the causal separation
of the measurements, one has to conclude that there must exist, locally
in the right wing, some elements of reality which pre-determine the
measurement outcome in the right wing. Einstein, Podolsky, and Rosen
formulated this idea in their famous \emph{Reality Criterion}:

\begin{quote}
If, without in any way disturbing a system, we can predict with certainty
(i.e., with probability equal to unity) the value of a physical quantity,
then there exists an element of reality corresponding to that quantity.
(\cite{key-6} p. 777) 
\end{quote}
It is probably true that no physicist would find this thesis implausible.
In our example, the value of the spin of the right particle in direction
$\mathbf{b}$ can be predicted with 100\% certainty by performing
a far distant spin measurement on the left particle in direction $\mathbf{b}$,
that is without in any way disturbing the right particle. Consequently,
there must exist some element of reality \emph{in the right wing},
that corresponds to the value of the spin of the right particle in
direction $\mathbf{b}$, in other words, there must exist something
in the right wing that determines the outcome of the spin measurement
on the right particle. 

One might think that if this is true for a given direction $\mathbf{b}$
then---by the same token---it must be true for all possible directions.
However, this is not necessarily the case. This is true only if the
following condition is satisfied: 

\medskip{}

\begin{center}
\begin{minipage}[c][1\totalheight]{0.9\columnwidth}%

\paragraph*{Assumption~4}

The choices between the measurement setups in the left and right wings
are entirely autonomous, that is, they are independent of each other
and of the assumed elements of reality that determine the measurement
outcomes. %
\end{minipage}%

\par\end{center}

\medskip{}

\noindent Otherwise the following conspiracy is possible: something
in the world pre-determines which measurement will be performed and
what will be the outcome. We assume however that there is no such
a conspiracy in our world. 

Thus, taking into account Assumptions~2, 3 and 4, we arrive at the
conclusion that there are elements of reality corresponding to the
values of the spin of the particles in all directions. (Of course,
it does not mean that we are able to predict the spin of the right
particle in all directions simultaneously. The reason is that we are
not able to measure the spin of the left particle in all directions
simultaneously.)

\subsection{Does quantum mechanics describe these elements of reality? }

The answer is no. However, the meaning of this {}``no'' is more
complex and depends on the interpretation of wave function (pure state). 

The \emph{Copenhagen interpretation} asserts that a pure state $\psi$
provides a complete and exhaustive description of an individual system,
and a dynamical variable represented by the operator $\hat{A}$ has
value $a$ if and only if $\hat{A}\psi=a\psi$. Consequently, spin
has a given value only if the state of the system is the corresponding
eigenvector of the spin-operator. But  spin-operators in different
directions do not commute, therefore there is no state in which spin
would have values in all directions. Thus, in fact, the EPR argument
must be considered as a strong argument against the Copenhagen interpretation
of wave function. 

According to the \emph{statistical interpretation}, a wave function
does not provide a complete description of an individual system but
only characterizes the system in a statistical/probabilistic sense.
The wave function is not tracing the complete ontology of the system.
Therefore, from the point of view of the statistical interpretation,
the novelty of the EPR argument consists in not proving that quantum
mechanics is incomplete but pointing out concrete elements of reality
that are outside of the scope of a quantum mechanical description.

It does not mean, however, that statistical interpretation remains
entirely untouched by the EPR argument. In fact the statistical interpretation
of quantum mechanics, as a probabilistic model in general, admits
different ontological pictures. And the EPR argument provides restrictions
for the possible ontologies. Consider the following simple example.
Imagine that we pull a die from a hat and throw it (event $D$). There
are six possible outcomes: $<1>,<2>,\ldots<6>$. By repeating this
experiment many times, we observe the following relative frequencies:\begin{equation}
\begin{array}{rcl}
p\left(<1>|D\right) & = & 0.05\\
p\left(<2>|D\right) & = & 0.1\\
p\left(<3>|D\right) & = & 0.1\\
p\left(<4>|D\right) & = & 0.1\\
p\left(<5>|D\right) & = & 0.1\\
p\left(<6>|D\right) & = & 0.55\end{array}\label{eq_kocka_sulyok}\end{equation}
$p\left(D\right)=1$, therefore $p\left(<1>\right)=0.05$,... $p\left(<6>\right)=0.55$.
Our probabilistic model will be based on these probabilities, and
it works well. It correctly describes the behavior of the system:
it correctly reflects the relative frequencies, correctly predicts
that the mean value of the thrown numbers is $4.75$, etc. In other
words, our probabilistic model provides everything expected from a
probabilistic model. However, there can be two different ontological
pictures behind this probabilistic description:

\begin{itemize}
\item [(A)] The dice in the hat are biased differently. Moreover, each
of them is biased by so much, the mass distribution is asymmetric
by so much, that practically (with probability $1$) only one outcome
is possible when we throw it. The distribution of the differently
biased dice in the hat is the following: 5\% of them are predestinated
for $<1>$, 10\% for $<2>$, ... and 55\% for $<6>$. That is to say,
each die in the hat has a pre-established \emph{property} (characterizing
its mass distribution). The dice throw---as a measurement---reveals
these properties. When we obtain result $<2>$, it reveals that the
die has property `2'. In other words, there exists a real event in
the world, namely

\begin{center}
$\widetilde{<2>}$ = the die we have just pulled from the hat has
property `2'
\par\end{center}

such that \begin{eqnarray}
p\left(\widetilde{<2>}\right) & = & p\left(<2>|D\right)\label{eq_kocka_tul1}\\
p\left(<2>|\widetilde{<2>}\right) & = & 1\label{eq_kocka_tul2}\end{eqnarray}
That is, in our example, event $\widetilde{<2>}$ occurs with probability
0.1 independently of whether we perform the dice throw or not. 

\item [(B)] All dice in the hat are uniformly prepared. Each of them has
the same slightly asymmetric mass distribution such that the outcome
of the throw can be anything with probabilities (\ref{eq_kocka_sulyok}).
In this case, if the result of the throw is $<2>$, say, it is meaningless
to say that the measurement revealed that the die has property `2'.
For the outcome of an individual throw tells nothing about the properties
of an individual die. In this case, there does not exist a real event
$\widetilde{<2>}$ for which (\ref{eq_kocka_tul1}) and (\ref{eq_kocka_tul2})
hold.

By repeating the experiment many times, we obtain the conditional
probabilities (\ref{eq_kocka_sulyok}). These conditional probabilities
collectively, that is, the conditional probability distribution over
all possible outcomes, do reflect an objective property common to
all individual dice in the hat, namely their mass distribution. (One
might think that (A) is a hidden variable interpretation of the probabilistic
model in question, while the situation described in (B) does not admit
a hidden variable explanation. It is entirely possible, however, that
events $\widetilde{<1>},\widetilde{<2>},\ldots$ are objectively indeterministic.
On the other hand, in case (B), the physical process during the dice
throw can be completely deterministic and the probabilities in question
can be epistemic.) 

\end{itemize}
We have a completely similar situation in quantum mechanics. Consider
an observable with a spectral decomposition $\hat{A}=\sum_{i}a_{i}P_{i}$.
It is not entirely clear what we mean by saying that {}``$tr\left(\hat{W}P_{i}\right)$
is the probability of that physical quantity $A$ has value $a_{i}$,
if the sate of the system is $\hat{W}$.'' To clarify the precise
meaning of this statement, let us start with what seems to be certain.
We assumed (Assumption~1) that the quantity $tr\left(\hat{W}P_{i}\right)$
is identified with the observed conditional probability $p\left(<a_{i}>|a\right)$,
where $a$ denotes the event consisting in the performing the measurement
itself and $<a_{i}>$ denotes the outcome event corresponding to pointer
position `$a_{i}$': \begin{equation}
tr\left(\hat{W}P_{i}\right)=p\left(<a_{i}>|a\right)\label{eq_mi_trace_min}\end{equation}
If nothing more is assumed, then a measurement outcome becomes fixed
during the measurement itself, and we obtain a type (B) interpretation
of quantum probabilities. Let us call this the \emph{minimal interpretation.}
In this case, a measurement outcome $<a_{i}>$ does not reveal a property
of the individual object. Of course, the state of the system, $\hat{W}$,
no matter whether it is a pure state or not, may reflect a property
of the individual objects, just like the conditional probabilities
(\ref{eq_kocka_sulyok}) reflect the mass distribution of the individual
dice. 

One can also imagine a type (A) interpretation of $tr\left(\hat{W}P_{i}\right)$,
which we call the \emph{property interpretation}. According to this
view, every individual measurement outcome $<a_{i}>$ corresponds
to an objective property $\widetilde{<a_{i}>}$ intrinsic to the individual
object, which is revealed by the measurement. This property exists
and is established independently of whether the measurement is performed
or not. Just as in the example above, equation (\ref{eq_mi_trace_min})
can be continued in the following way:\begin{equation}
tr\left(\hat{W}P_{i}\right)=p\left(<a_{i}>|a\right)=p\left(\widetilde{<a_{i}>}\right)\label{eq_mi_trace_tul}\end{equation}
where $p\left(\widetilde{<a_{i}>}\right)$ is the probability of that
the individual object in question has the property $\widetilde{<a_{i}>}$.

Now, from the EPR argument we conclude that the ontological picture
provided by the type (B) interpretation is not satisfactory. For according
to the EPR argument there must exist previously established elements
of reality that determine the outcomes of the individual measurements.
This claim is nothing but a type (A) interpretation.

\subsection{The EPR conclusion\label{sub:The-EPR-conclusion}}

One has to emphasize that the conclusion of the EPR argument is not
a no-go theorem for hidden variable models of quantum mechanics. On
the contrary, it asserts that \emph{there must be a more complete
description of physical reality behind quantum mechanics}. There must
be a state, a hidden variable, characterizing the state of affairs
in the world in more detail than the quantum mechanical state operator,
\emph{something that also reflects the missing elements of reality}.
In other words, the pre-established value of the hidden variable has
to determine the spin of both particles in all possible directions.
Perhaps it is not fair to quote Einstein himself in this context,
who was not completely satisfied with the published version of the
joint paper (see \cite{key-8}), but in this final conclusion there
seems to be an agreement: 

\begin{quote}
I am, in fact, firmly convinced that the essentially statistical character
of contemporary quantum theory is solely to be ascribed to the fact
that this theory operates with an incomplete description of physical
systems. (Quoted by \cite{key-9}, p. 90.) 
\end{quote}
Also, the EPR paper ended with:

\begin{quote}
While we have thus shown that the wave function does not provide a
complete description of the physical reality, we left open the question
of whether or not such a description exists. We believe, however,
that such a theory is possible.
\end{quote}
The question is: do these missing elements of reality really exist?
We will answer this question in section~\ref{sec:Do-the-missing}
after some technical preparations.

\section{Under what conditions can a system of empirically ascertained probabilities
be described by Kolmogorov's probability theory?}

The following mathematical preparations will provide some probability
theoretic inequalities which are not identical with but deeply related
to the Bell-type inequalities; they play an important role in distinguishing
classical Kolmogorovian probabilities from quantum probabilities.

\subsection{Pitowsky theorem\label{sub:Pitowsky-theorem}}

Imagine that somehow we assign numbers between $0$ and $1$ to particular
events, and we regard them as {}``probabilities'' in some intuitive
sense. Under what conditions can these {}``probabilities'' be represented
in a Kolmogorovian probabilistic theory? As we will see, such a representation
is always possible. Restrictive conditions will be obtained only if
we also want to represent some of the correlations among the events
in question. 

Consider the following events: $A_{1},A_{2},\ldots A_{n}$. Let\[
S\subseteq\left\{ (i,j)|i<j;i,j=1,2,\dots n\right\} \]
be a set of pairs of indexes corresponding to those pairs of events
the correlations of which we want to be represented. The following
{}``probabilities'' are given:\begin{equation}
\begin{array}{rclcl}
p_{i} & = & p\left(A_{i}\right) &  & i=1,2,\dots n\\
p_{ij} & = & p\left(A_{i}\wedge A_{j}\right) &  & (i,j)\in S\end{array}\label{eq_empirikus_valoszinusegek}\end{equation}
 We say that {}``probabilities'' (\ref{eq_empirikus_valoszinusegek})
have Kolmogorovian representation if there is a Kolmogorovian probability
model $\left(\Sigma,\mu\right)$ with some $X_{1},X_{2},\dots X_{n}\in\Sigma$
elements of the event algebra, such that\begin{equation}
\begin{array}{rclcl}
p_{i} & = & \mu\left(X_{i}\right) &  & i=1,2,\dots n\\
p_{ij} & = & \mu\left(X_{i}\wedge X_{j}\right) &  & (i,j)\in S\end{array}\label{eq_kolmogorovi_reprezentacio}\end{equation}

The question is, under what conditions does there exist such a representation?
It is interesting that this evident problem was not investigated until
the pioneer works of Accardi \cite{key-1,key-3} and Pitowsky \cite{key-4}
in the late 80's. 

For the discussion of the problem, Pitowsky introduced an expressive
geometric language. From the probabilities (\ref{eq_empirikus_valoszinusegek})
we compose an $n+\left|S\right|$-dimensional, so called, \emph{correlation
vector} ($\left|S\right|$ denotes the cardinality of $S$):\[
\overrightarrow{p}=\left(p_{1},p_{2},\ldots p_{n},\ldots p_{ij},\ldots\right)\]
 Denote $R(n,S)\cong\mathbb{R}^{n+\left|S\right|}$ the linear space
consisting of real vectors of this type. Let $\varepsilon\in\left\{ 0,1\right\} ^{n}$
be an arbitrary $n$-dimensional vector consisting of $0$'s and $1$'s.
For each $\varepsilon$ we construct the following $\overrightarrow{u}^{\varepsilon}\in R(n,S)$
vector: \begin{equation}
\begin{array}{rclcl}
u_{i}^{\varepsilon} & = & \varepsilon_{i} &  & i=1,2,\ldots n\\
u_{ij}^{\varepsilon} & = & \varepsilon_{i}\varepsilon_{j} &  & (i,j)\in S\end{array}\label{eq:vertex}\end{equation}
The set of convex linear combinations of $u^{\varepsilon}$'s is called
a classical correlation polytope:\[
c(n,S)=\left\{ \overrightarrow{f}\in R(n,S)\left|\overrightarrow{f}=\sum_{\varepsilon\in\left\{ 0,1\right\} ^{n}}\lambda_{\varepsilon}\overrightarrow{u}^{\varepsilon}\right.;\,\lambda_{\varepsilon}\geq0;\,\sum_{\varepsilon\in\left\{ 0,1\right\} ^{n}}\lambda_{\varepsilon}=1\right\} \]

In 1989, Pitowsky proved (\cite{key-4}, pp. 22--24) the following
theorem:

\paragraph*{Theorem}

\emph{The correlation vector} $\overrightarrow{p}$ \emph{admits a
Kolmogorovian representation if and only if} $\overrightarrow{p}\in c(n,S)$.
\medskip{}

Beyond the fact that the theorem plays an important technical role
in the discussions of the EPR--Bell problem and other foundational
questions of quantum theory, it shades light on an interesting relationship
between classical propositional logic and Kolmogorovian probability
theory. We must recognize that the vertices of $c(n,S)$ defined in
(\ref{eq:vertex}) are nothing but the classical two-valued truth-value
functions over a minimal propositional algebra naturally related to
events $A_{1},A_{2},\ldots A_{n}$. Therefore, what the theorem says
is that probability distributions are nothing but weighted averages
of the classical truth-value functions.

\subsection{Inequalities}

It is a well known mathematical fact that the conditions for a vector
to fall into a convex polytope can be expressed by a set of linear
inequalities. What kind of inequalities express the condition $\overrightarrow{p}\in c(n,S)$? 

\begin{figure}[htbp]
\begin{centering}
\includegraphics[scale=0.5]{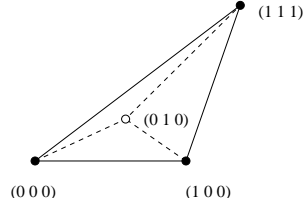}
\par\end{centering}

\caption{\emph{In the case of} $n=2$, \emph{classical correlation polytope
has four vertices \label{fig_n=00003D2}}}

\end{figure}

The answer is trivial in the case of $n=2$ and $S=\left\{ (1,2)\right\} $.
Set $\left\{ 0,1\right\} ^{2}$ has four elements: $(0,0)$, $(1,0)$,
$(0,1)$, and $(1,1).$ Consequently the classical correlation polytope
(Fig.~\ref{fig_n=00003D2}) has four vertices: $(0,0,0)$, $(1,0,0)$,
$(0,1,0)$, and $(1,1,1)$. 

The condition $\overrightarrow{p}\in c(2,S)$ is equivalent with the
following inequalities:\begin{equation}
\begin{array}{c}
0\leq p_{12}\leq p_{1}\leq1\\
0\leq p_{12}\leq p_{2}\leq1\\
p_{1}+p_{2}-p_{12}\leq1\end{array}\label{eq_negyenlo2}\end{equation}
 Indeed, from (\ref{eq_negyenlo2}) we have:\begin{eqnarray*}
\overrightarrow{p} & = & \left(1-p_{1}-p_{2}+p_{12}\right)\left(\begin{array}{c}
0\\
0\\
0\end{array}\right)+\left(p_{1}-p_{12}\right)\left(\begin{array}{c}
1\\
0\\
0\end{array}\right)\\
 &  & +\left(p_{2}-p_{12}\right)\left(\begin{array}{c}
0\\
1\\
0\end{array}\right)+p_{12}\left(\begin{array}{c}
1\\
1\\
1\end{array}\right)\end{eqnarray*}

Another important case is when $n=3$ and $S=\left\{ (1,2),(1,3),(2,3)\right\} .$
The corresponding set of inequalities is the following (\cite{key-4},
pp. 25--26):\begin{equation}
\begin{array}{c}
0\leq p_{ij}\leq p_{i}\leq1\\
0\leq p_{ij}\leq p_{j}\leq1\\
p_{i}+p_{j}-p_{ij}\leq1\\
p_{1}+p_{2}+p_{3}-p_{12}-p_{13}-p_{23}\leq1\\
p_{1}-p_{12}-p_{13}+p_{23}\geq0\\
p_{2}-p_{12}-p_{23}+p_{13}\geq0\\
p_{3}-p_{13}-p_{23}+p_{12}\geq0\end{array}\label{eq_Bell-Pitowsky}\end{equation}
These are the \emph{Bell--Pitowsky inequalities}.

Finally we mention the case of $n=4$ and \[
S=\left\{ (1,3),(1,4),(2,3),(2,4)\right\} \]
One can prove (\cite{key-4}, pp. 27--30) that the following inequalities
are equivalent with the condition $\overrightarrow{p}\in c(4,S)$:
\begin{equation}
\begin{array}{ccc}
0\leq p_{ij}\leq p_{i}\leq1\\
0\leq p_{ij}\leq p_{j}\leq1 &  & i=1,2\,\,\, j=3,4\\
p_{i}+p_{j}-p_{ij}\leq1\\
\begin{array}{ccccc}
-1 & \leq & p_{13}+p_{14}+p_{24}-p_{23}-p_{1}-p_{4} & \leq & 0\\
-1 & \leq & p_{23}+p_{24}+p_{14}-p_{13}-p_{2}-p_{4} & \leq & 0\\
-1 & \leq & p_{14}+p_{13}+p_{23}-p_{24}-p_{1}-p_{3} & \leq & 0\\
-1 & \leq & p_{24}+p_{23}+p_{13}-p_{14}-p_{2}-p_{3} & \leq & 0\end{array}\end{array}\label{eq_CH-Pitowsky}\end{equation}
Let us call them the \emph{Clauser--Horne--Pitowsky inequalities}.

\section{Do the missing elements of reality exist?\label{sec:Do-the-missing}}

The elements of reality the EPR paper is talking about are nothing
but what the property interpretation calls properties existing independently
of the measurements. In each run of the experiment, there exist some
elements of reality, the system has particular properties $\widetilde{<a_{i}>}$
which unambiguously determine the measurement outcome $<a_{i}>$,
\emph{given} that the corresponding measurement $a$ is performed.
That is to say, \begin{equation}
p\left(<a_{i}>|\widetilde{<a_{i}>}\wedge a\right)=1\label{eq:tul-determinal}\end{equation}
(This condition---coming from Assumptions~2 and 3 and the Reality
Criterion---is sometimes called {}``Counterfactual Definiteness''
\cite{key-55}.) According to the {}``no conspiracy'' assumption
we stipulated in Assumption~4, \begin{equation}
p\left(\widetilde{<a_{i}>}\wedge a\right)=p\left(\widetilde{<a_{i}>}\right)p\left(a\right)\label{eq:no-conspiracy}\end{equation}
so (\ref{eq:tul-determinal}) and (\ref{eq:no-conspiracy}) imply
that\begin{equation}
p\left(\widetilde{<a_{i}>}\right)=p\left(<a_{i}>|a\right)=tr\left(\hat{W}P_{i}\right)\label{eq:kettos}\end{equation}
That is, the relative frequency of the element of reality $\widetilde{<a_{i}>}$
corresponding to the measurement outcome $<a_{i}>$ must be equal
to the corresponding quantum probability $tr\left(\hat{W}P_{i}\right)$.
However, this is generally impossible. According to the \emph{Laboratory
Record Argument} \cite{key-10} below, there are no things (elements
of reality, properties, {}``quantum events'', etc.) the relative
frequencies of which could be equal to quantum probabilities.

\begin{figure}
\begin{centering}
\includegraphics[width=0.45\columnwidth,keepaspectratio]{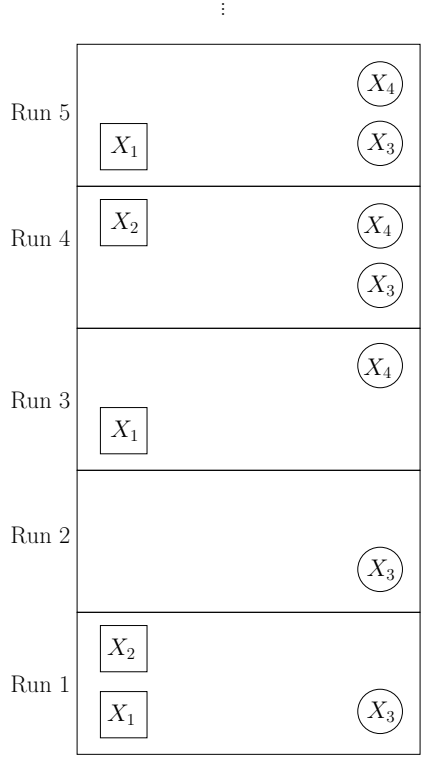}
\par\end{centering}

\caption{\emph{In each run of the experiment, some of the things in question
(elements of reality, properties, {}``quantum events'', etc.) occur}\label{fig:In-each-run}}

\end{figure}

Imagine the consecutive time slices of a given region of the world
(say, the laboratory) corresponding to the consecutive runs of an
experiment (Fig.~\ref{fig:In-each-run}). We do not know what {}``elements
of reality'', {}``properties'', {}``quantum events'', etc., are,
but we can imagine that in every such time slices some of them occur,
and we can imagine a %
 laboratory record like the one in Table~\ref{tab_labor_konyv}.%
\begin{table}[htbp]
\begin{centering}
\begin{tabular}{|c|c|c|c|c|c|c|c|c|}
\hline 
Run & {\small $X_{1}$} & {\small $X_{2}$} & {\small $X_{3}$} & {\small $X_{4}$} & {\small $X_{1}\wedge X_{3}$} & {\small $X_{1}\wedge X_{4}$} & {\small $X_{2}\wedge X_{3}$} & {\small $X_{2}\wedge X_{4}$}\tabularnewline
\hline
\hline 
{\small 1} & {\small 1} & {\small 1} & {\small 1} & {\small 0} & {\small 1} & {\small 0} & {\small 1} & {\small 0}\tabularnewline
\hline 
{\small 2} & {\small 0} & {\small 0} & {\small 1} & {\small 0} & {\small 0} & {\small 0} & {\small 0} & {\small 0}\tabularnewline
\hline 
{\small 3} & {\small 1} & {\small 0} & {\small 0} & {\small 1} & {\small 0} & {\small 1} & {\small 0} & {\small 0}\tabularnewline
\hline 
{\small 4} & {\small 0} & {\small 1} & {\small 1} & {\small 1} & {\small 0} & {\small 0} & {\small 1} & {\small 1}\tabularnewline
\hline 
{\small 5} & {\small 1} & {\small 0} & {\small 0} & {\small 0} & {\small 0} & {\small 0} & {\small 0} & {\small 0}\tabularnewline
\hline 
{\small 6} & {\small 0} & {\small 1} & {\small 0} & {\small 1} & {\small 0} & {\small 0} & {\small 0} & {\small 1}\tabularnewline
\hline 
{\small 7} & {\small 0} & {\small 1} & {\small 0} & {\small 1} & {\small 0} & {\small 0} & {\small 0} & {\small 1}\tabularnewline
\hline 
{\small 8} & {\small 1} & {\small 0} & {\small 0} & {\small 1} & {\small 0} & {\small 1} & {\small 0} & {\small 0}\tabularnewline
\hline 
{\small $\vdots$} & {\small $\vdots$} & {\small $\vdots$} & {\small $\vdots$} & {\small $\vdots$} & {\small $\vdots$} & {\small $\vdots$} & {\small $\vdots$} & {\small $\vdots$}\tabularnewline
\hline 
{\small 99998} & {\small 1} & {\small 0} & {\small 0} & {\small 0} & {\small 0} & {\small 0} & {\small 0} & {\small 0}\tabularnewline
\hline 
{\small 99999} & {\small 0} & {\small 0} & {\small 1} & {\small 0} & {\small 0} & {\small 0} & {\small 0} & {\small 0}\tabularnewline
\hline 
{\small $N$=100000} & {\small 0} & {\small 1} & {\small 0} & {\small 1} & {\small 0} & {\small 0} & {\small 0} & {\small 1}\tabularnewline
\hline 
 & {\small $N_{1}$} & {\small $N_{2}$} & {\small $N_{3}$} & {\small $N_{4}$} & {\small $N_{13}$} & {\small $N_{14}$} & {\small $N_{23}$} & {\small $N_{24}$}\tabularnewline
\hline
\end{tabular}
\par\end{centering}

\caption{\emph{An imaginary laboratory record about the occurrences of the
hidden elements of reality\label{tab_labor_konyv}}}

\end{table}
`1' stands for the case if the corresponding element of reality occurs
and `0' if it does not. We put `1' into the column corresponding to
a conjunction if both elements of reality occur. In order to avoid
the objections like {}``the two measurements cannot be performed
simultaneously'', or {}``the conjunction is meaningless'', etc.,
let us assume that the pairs $\left(X_{1},X_{3}\right)$, $\left(X_{1},X_{4}\right)$,
$\left(X_{2},X_{3}\right)$, and $\left(X_{2},X_{4}\right)$ belong
to commuting projectors. 

Now, the relative frequencies can be computed from this table: \begin{equation}
\nu_{1}=\frac{N_{1}}{N},\nu_{1}=\frac{N_{2}}{N},\ldots\nu_{24}=\frac{N_{24}}{N}\label{eq_jegyzokonyv_rel_gyak}\end{equation}
Notice that each row of the table corresponds to one of the $2^{4}$
possible classical truth-value functions over the corresponding propositions.
In other words, it is one of the vertices $\overrightarrow{u}^{\varepsilon}$
($\varepsilon\in\left\{ 0,1\right\} ^{4}$) we introduced in (\ref{eq:vertex}).
Let $N_{\varepsilon}$ denote the number of type-$\overrightarrow{u}^{\varepsilon}$
rows in the table. The relative frequencies (\ref{eq_jegyzokonyv_rel_gyak})
can also be expressed as follows:\begin{eqnarray*}
\nu_{i} & = & \sum_{\varepsilon\in\left\{ 0,1\right\} ^{4}}\lambda_{\varepsilon}u_{i}^{\varepsilon}\\
\nu_{ij} & = & \sum_{\varepsilon\in\left\{ 0,1\right\} ^{4}}\lambda_{\varepsilon}u_{ij}^{\varepsilon}\end{eqnarray*}
where $\lambda_{\varepsilon}=\frac{N_{\varepsilon}}{N}$. Clearly,
$\lambda_{\varepsilon}\geq0$ and $\sum_{\varepsilon\in\left\{ 0,1\right\} ^{4}}\lambda_{\varepsilon}=1$.
That is to say, the correlation vector consisting of the relative
frequencies in question satisfies the condition $\overrightarrow{\nu}=\left(\nu_{1},\nu_{2},\ldots\nu_{24}\right)\in c(4,S)$
in section~\ref{sub:Pitowsky-theorem}. (Consequently---due to Pitowsky's
theorem---it admits a Kolmogorovian representation.) 

One can generalize the above observation in the following stipulation:
The elements of a correlation vector $\overrightarrow{p}$ admit a
relative frequency interpretation if and only if $\overrightarrow{p}$
satisfies the condition $\overrightarrow{p}\in c(n,S)$.

So in the above example, $\overrightarrow{\nu}\in c(4,S)$ if and
only if $\overrightarrow{\nu}$ satisfies the Clauser--Horne--Pitowsky
inequalities (\ref{eq_CH-Pitowsky}). But, in general, quantum probabilities
do not satisfy these inequalities. Consider the EPR experiment in
section~\ref{sub:The-description-of}. Assume that the possible directions
are $\mathbf{a}_{1}$ and $\mathbf{a}_{2}$ in the left wing, and
$\mathbf{b}_{1}$ and $\mathbf{b}_{2}$ in the right wing. We will
consider the following particular case: $\sphericalangle\left(\mathbf{a}_{1}\mathbf{,b}_{1}\right)=\sphericalangle\left(\mathbf{a}_{1}\mathbf{,b}_{2}\right)=\sphericalangle\left(\mathbf{a}_{2}\mathbf{,b}_{2}\right)=120^{\circ}$
and $\sphericalangle\left(\mathbf{a}_{2}\mathbf{,b}_{1}\right)=0$.
According to (\ref{eq_EPR_prob_fel})--(\ref{eq_EPR_prob_sin}), the
quantum probabilities are the following:\begin{eqnarray}
p(A_{1}|a_{1})=p(A_{2}|a_{2})=p(B_{1}|b_{1})=p(B_{2}|b_{2}) & = & \frac{1}{2}\label{eq_EPR_szamok1}\\
p(A_{1}\wedge B_{1}|a_{1}\wedge b_{1})=p(A_{1}\wedge B_{2}|a_{1}\wedge b_{2})\nonumber \\
=p(A_{2}\wedge B_{2}|a_{2}\wedge b_{2}) & = & \frac{3}{8}\label{eq_EPR_szamok2}\\
p(A_{2}\wedge B_{1}|a_{2}\wedge_{1}b) & = & 0\label{eq_EPR_szamok3}\end{eqnarray}
Let $X_{1}=A_{1},X_{2}=A_{2},X_{3}=B_{1}$, and $X_{4}=B_{2}$. The
question is whether the corresponding correlation vector $\overrightarrow{p}=\left(\frac{1}{2},\frac{1}{2},\frac{1}{2},\frac{1}{2},\frac{3}{8},\frac{3}{8},0,\frac{3}{8}\right)$
satisfies the condition of Kolmogorovity or not. Substituting the
elements of $\overrightarrow{p}$ into (\ref{eq_CH-Pitowsky}), we
find that the system of inequalities is violated. Quantum probabilities
measured in the EPR experiment violate the Clauser--Horne--Pitowsky
inequalities, therefore they cannot be interpreted as relative frequencies.
Consequently, \emph{there cannot exist quantum events, elements of
reality, properties, or any other things which occur with relative
frequencies equal to quantum probabilities.} (To avoid any misunderstanding,
the \emph{restriction} of a quantum probability measure to the Boolean
sublattice of projectors belonging to the spectral decomposition of
one single maximal observable does, of course, admit a relative frequency
interpretation. It must be also mentioned that quantum probabilities,
in general, can be interpreted in terms of relative frequencies as
\emph{conditional} probabilities \cite{key-10}.) 

In brief, \emph{given the existence of the predicted perfect correlations
by quantum mechanics (Assumption~3), according to the EPR argument,
there ought to exist particular elements of reality, which, according
to the Laboratory Record Argument, cannot exist. To resolve this contradiction,
we have to conclude that at least one of Assumption~1, 2 and 4 fails}. 

In the next section we will arrive at similar conclusions in a different
context.

\section{Bell's inequalities\index{Bell-tétel}\label{sec:Bell-t=0000E9tel}}

\subsection{Bell's formulation of the problem}

When the EPR paper was published, there already existed a hidden variable
theory of quantum mechanics, which achieved its complete form in 1952
\cite{key-23,key-24}. This is the de Broglie--Bohm theory, which
also called Bohmian mechanics. (For a historical review of the de
Broglie--Bohm theory, see \cite{key-25}. For the Bohmian mechanics
version of the standard text-book quantum mechanics, see \cite{key-26}
and \cite{key-27}.) This theory is explicitly non-local in the following
sense: One of its central objects, the so called quantum potential
which locally governs the behavior of a particle, explicitly depends
on the simultaneous coordinates of other, far distant, particles.
This kind of non-locality is, however, a natural feature of all theories
containing potentials (like electrostatics or the Newtonian theory
of gravitation). Such a theory is expected to describe physical reality
only in non-relativistic approximation, when the finiteness of the
speed of propagation of causal effects is negligible, but, according
to our expectations, it fails on a more detailed spatiotemporal scale.
What is unusual in the EPR situation is that the real laboratory experiments
do reach this relativistic spatiotemporal scale, but the observed
results are still describable by simple (non-local) quantum/Bohm mechanics. 

In his 1964 paper \cite{key-29,key-9}, John Stuart\index{Bell} Bell
proved that 

\begin{quote}
In a theory in which parameters are added to quantum mechanics to
determine the results of individual measurements, without changing
the statistical predictions, there must be a mechanism whereby the
setting of one measuring device can influence the reading of another
instrument, however remote. (\cite{key-9}, p. 20.) 
\end{quote}
The argument was based on the violation of an inequality derivable
from a few plausible assumptions. Instead of Bell's original inequality,
it is better to formulate the argument by means of the Clauser--Horne
inequalities, which are more applicable to the spin-correlation experiment
described in section~\ref{sub:The-description-of}. This difference
is, however, not significant. 

Bell was concerned with the following problem: Can the whole EPR experiment
be accommodated in a classical world, that is, in a world which is
compatible with the world-view of pre-quantum-mechanical physics?
This pre-quantum-mechanical world is local, deterministic and Markovian
(LDM), that is, it satisfies the following assumption:%
\begin{figure}
\begin{centering}
\includegraphics[width=0.5\columnwidth]{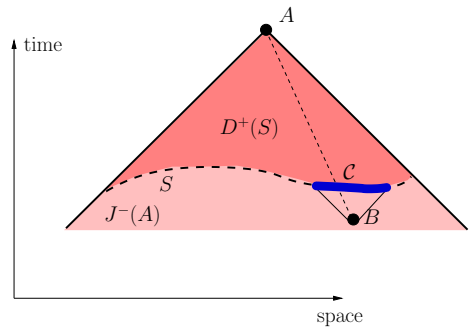}
\par\end{centering}

\caption{\emph{A local, deterministic, and Markovian (LDM) world. Event} $A$
\emph{is determined by the history of the universe inside of the backward
light-cone} $J^{-}(A)$\emph{. The state of affairs along a Cauchy
hyper-surface} $S$ \emph{completely determines the history within
the dependence domain} $D^{+}(S)$\emph{. (For these basic concepts
of relativity theory, see \cite{key-30,key-31}.) In other words,
all the relevant information from the past is encoded in the state
of affairs in the present. More exactly, all information from a past
event} $B$ \emph{influencing} $A$ \emph{must be encoded in the corresponding
region} $\mathcal{C}$ \emph{\label{cap:A-local-deterministic}}}

\end{figure}

\medskip{}

\begin{center}
\begin{minipage}[c][1\totalheight]{0.9\columnwidth}%

\paragraph*{Assumption~2'}

Our world is 

\begin{enumerate}
\item Local---No direct causal connection between spatially separated events
(Assumption~2).
\item Deterministic---Event $A$ is uniquely determined by the pre-history
in the backward light-cone $J^{-}(A)$. (Fig.~\ref{cap:A-local-deterministic})
\item Markovian---All the relevant information from the past is encoded
in the state of affairs in the present.
\end{enumerate}
\end{minipage}%

\par\end{center}

\medskip{}

\noindent Electrodynamics is the paradigmatic LDM theory of this pre-quantum-mechanical
world view. 

It should be clear that Assumption~2' prescribes determinism only
on the level of the final ontology, but it does not exclude stochasticity
of an epistemic kind. At first sight Assumption~2' seems to be much
stronger than Assumption~2. It is because the three metaphysical
ideas, locality, determinism, and Markovity, seem to be clearly distinguishable
features of a possible world. However, further reflection reveals
that these concepts are inextricably intertwined. In all pre-quantum-mechanical
examples the laws of physics are such that locality, determinism,
and Markovity are provided together. If, however, our world is objectively
indeterministic---this, of course, hinges on the very issue we are
discussing here---then it is far from obvious how the phrase {}``no
direct causal connection between $\dots$'' is understood (also see
section~\ref{sec:No-correlation-without}). 

Anyhow, the question we are concerned with is this: \emph{Can all
physical events observed in the EPR experiment be accommodated in
an LDM world, including the emissions, the measurement setups, the
measurement outcomes, etc., with relative frequencies observed in
the laboratory and predicted by quantum mechanics? }

\subsection{The derivation of Bell's inequalities}

\begin{figure}[htbp]
\begin{centering}
\includegraphics[width=0.5\columnwidth]{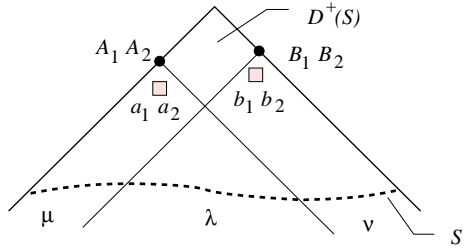}
\par\end{centering}

\caption{\emph{The space-time diagram of a single run of the EPR experiment\label{fig_EPR_elhelyezes}}}

\end{figure}

We have eight different types of event: the measurement outcomes,
that is, the detections of the particles in the corresponding up-detector,
$A_{1},A_{2},B_{1},B_{2}$, and the measurement setups $a_{1},a_{2},b_{1},b_{2}$.
Let us imagine the space-time diagram of one single run of the experiment
(Fig.~\ref{fig_EPR_elhelyezes}). %
 The positive dependence domain of the Cauchy surface $S$, $D^{+}(S)$,
contains all events we observe in a single run of the experiment.
According to the classical views, the Cauchy data on $S$ unambiguously
determine what is going on in domain $D^{+}(S)$, including whether
or not events $A_{1},A_{2},B_{1},B_{2},a_{1},a_{2},b_{1}$, and $b_{2}$
occur. The occurrence of a type-$X$ event means that the state of
affairs in the dependence domain $D^{+}(S)$ falls into the category
$X$. Which events occur and which do not, can be expressed with the
following functions:\begin{equation}
u^{X}\left(\mu,\lambda,\nu\right)=\left\{ \begin{array}{ll}
1 & \textrm{if }D^{+}(S)\textrm{ falls into category }X\\
0 & \textrm{if not}\end{array}\right.\label{eq_u_fuggveny_def_elso}\end{equation}
Taking into account that an event cannot depend on data outside of
the backward light-cone, \begin{equation}
\begin{array}{rcl}
u^{A_{i}}\left(\mu,\lambda,\nu\right) & = & u^{A_{i}}\left(\mu,\lambda\right)\\
u^{B_{i}}\left(\mu,\lambda,\nu\right) & = & u^{B_{i}}\left(\lambda,\nu\right)\\
u^{a_{i}}\left(\mu,\lambda,\nu\right) & = & u^{a_{i}}\left(\mu,\lambda\right)\\
u^{b_{i}}\left(\mu,\lambda,\nu\right) & = & u^{b_{i}}\left(\lambda,\nu\right)\end{array}\,\,\,\, i=1,2\label{eq_u_fuggvenyek}\end{equation}
\begin{figure}[h]
\begin{centering}
\includegraphics[width=0.4\columnwidth]{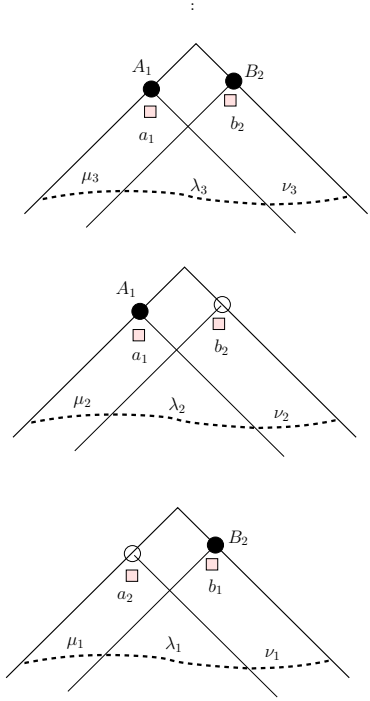}
\par\end{centering}

\caption{\emph{The statistical ensemble consists of the consecutive repetitions
of space-time pattern in Fig.~\ref{fig_EPR_elhelyezes} \label{cap:The-statistical-ensemble}}}

\end{figure}

The whole experiment, that is the statistical ensemble consist of
a long sequence of similar space-time patterns like the one depicted
in Fig.~\ref{fig_EPR_elhelyezes}. In the consecutive situations,
the existing values of parameters $\left(\mu,\lambda,\nu\right)$
determine what happens in the given run of the experiment (Fig.~\ref{cap:The-statistical-ensemble}).
One can count the relative frequencies of the various $\left(\mu,\lambda,\nu\right)$
combinations. Therefore, probabilities $p\left(\mu\right),p\left(\lambda\right),p\left(\nu\right),p\left(\mu\wedge\lambda\right),\ldots p\left(\mu\wedge\lambda\wedge\nu\right)$
can be considered as given. Applying (\ref{eq_u_fuggvenyek}), the
probabilities (relative frequencies) of the eight events can be expressed
as follows:\begin{eqnarray}
p\left(A_{i}\right) & = & \sum_{\mu,\lambda}u^{A_{i}}\left(\mu,\lambda\right)p\left(\mu\wedge\lambda\right)\label{eq_szallas1}\\
p\left(B_{i}\right) & = & \sum_{\lambda,\nu}u^{B_{i}}\left(\lambda,\nu\right)p\left(\lambda\wedge\nu\right)\\
p\left(a_{i}\right) & = & \sum_{\mu,\lambda}u^{a_{i}}\left(\mu,\lambda\right)p\left(\mu\wedge\lambda\right)\\
p\left(b_{i}\right) & = & \sum_{\lambda,\nu}u^{b_{i}}\left(\lambda,\nu\right)p\left(\lambda\wedge\nu\right)\\
p\left(A_{i}\wedge B_{j}\right) & = & \sum_{\mu,\lambda,\nu}u^{A_{i}}\left(\mu,\lambda\right)u^{B_{j}}\left(\lambda,\nu\right)p\left(\mu\wedge\lambda\wedge\nu\right)\\
p\left(a_{i}\wedge b_{j}\right) & = & \sum_{\mu,\lambda,\nu}u^{a_{i}}\left(\mu,\lambda\right)u^{b_{j}}\left(\lambda,\nu\right)p\left(\mu\wedge\lambda\wedge\nu\right)\label{eq_szallas6}\end{eqnarray}
. 

\begin{figure}[htbp]
\begin{centering}
\includegraphics[width=0.7\columnwidth]{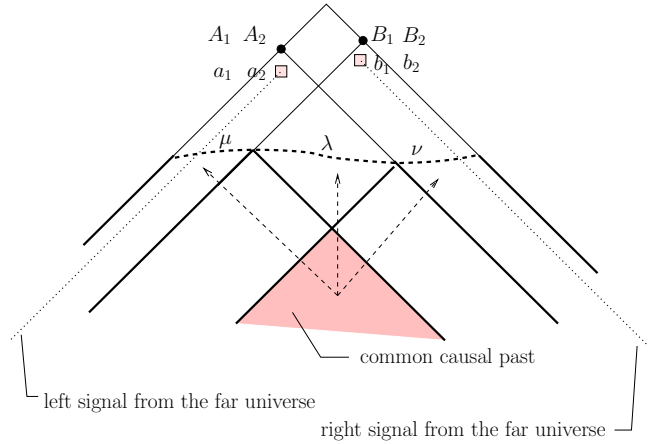}
\par\end{centering}

\caption{\emph{Due to the common causal past, there can be correlation between
the Cauchy data belonging to the three spatially separated regions.
One can, however, assume that the measurement setups are governed
by some independent signals coming from the far universe \label{fig_3fuggetlen}}}

\end{figure}

Due to the common causal past, there can be correlations between the
Cauchy data belonging to the three spatially separated regions (Fig.~\ref{fig_3fuggetlen}).
Henceforth, however, we assume that \begin{equation}
p\left(\mu\wedge\lambda\wedge\nu\right)=p\left(\mu\right)p\left(\lambda\right)p\left(\nu\right)\label{eq_harom_adat_fuggetlen2}\end{equation}
This assumption can be justified by the following intuitive arguments:

\begin{enumerate}
\item Our concern is to explain correlations between spatially separated
events observed in the EPR experiment. It would be completely pointless
to explain these correlation with similar correlations between earlier
spatially separated events. Because then we could say that a correlation
observed in a here-and-now experiment can be explained by something
around the Big Bang. 
\item In general, $\mu,\lambda$, and $\nu$ stand for huge numbers of Cauchy
data, depending on how detailed the description of the process in
question should be. Yet it is reasonable to assume that these parameters
only represent those data that are relevant for the events observed
in the EPR experiment. For example, one can imagine a scenario in
which the role of $\mu$ and $\nu$ is merely to govern the choice
of measurement setups in the left and in the right wing, and the values
of $\mu$ and $\nu$ are fixed by two independent assistants on the
left and right hand sides. In this case, it is quite plausible that
the free-will decisions of the assistants are independent of each
other, and also independent of parameter $\lambda$. 
\item If for any reason we do not like to appeal to free will, we can assume
that parameters $\mu$ and $\nu$, responsible for the measurement
setups, are determined by some random signals coming from the far
universe (Fig.~\ref{fig_3fuggetlen}). Also, we can assume that the
left and right signals are independent of each other and independent
of the value of $\lambda$---unless we want the explanation to go
back to the initial Big Bang singularity. 
\end{enumerate}
Applying Bayes rule and taking into account assumption (\ref{eq_harom_adat_fuggetlen2}),
the conditional probability $p\left(A_{i}\wedge B_{j}|a_{i}\wedge b_{j}\wedge\lambda\right)$
can be expressed as follows: \begin{eqnarray*}
 &  & \frac{p\left(A_{i}\wedge B_{j}\wedge a_{i}\wedge b_{j}\wedge\lambda\right)}{p\left(a_{i}\wedge b_{j}\wedge\lambda\right)}\\
 & = & \frac{\sum_{\mu,\nu}u^{A_{i}}\left(\mu,\lambda\right)u^{a_{i}}\left(\mu,\lambda\right)u^{B_{j}}\left(\lambda,\nu\right)u^{b_{j}}\left(\lambda,\nu\right)p\left(\mu\right)p\left(\nu\right)p\left(\lambda\right)}{\sum_{\mu,\nu}u^{a_{i}}\left(\mu,\lambda\right)u^{b_{j}}\left(\lambda,\nu\right)p\left(\mu\right)p\left(\nu\right)p\left(\lambda\right)}\\
 & = & \frac{\sum_{\mu}u^{A_{i}}\left(\mu,\lambda\right)p\left(\mu\right)p\left(\lambda\right)}{\sum_{\mu}u^{a_{i}}\left(\mu,\lambda\right)p\left(\mu\right)p\left(\lambda\right)}\,\frac{\sum_{\nu}u^{B_{j}}\left(\lambda,\nu\right)p\left(\nu\right)p\left(\lambda\right)}{\sum_{\nu}u^{b_{j}}\left(\lambda,\nu\right)p\left(\nu\right)p\left(\lambda\right)}\\
 & = & \frac{\sum_{\mu}u^{A_{i}}\left(\mu,\lambda\right)u^{a_{i}}\left(\mu,\lambda\right)p\left(\mu\right)p\left(\lambda\right)}{\sum_{\mu}u^{a_{i}}\left(\mu,\lambda\right)p\left(\mu\right)p\left(\lambda\right)}\\
 &  & \times\frac{\sum_{\nu}u^{B_{j}}\left(\lambda,\nu\right)u^{b_{j}}\left(\lambda,\nu\right)p\left(\nu\right)p\left(\lambda\right)}{\sum_{\nu}u^{b_{j}}\left(\lambda,\nu\right)p\left(\nu\right)p\left(\lambda\right)}\\
 & = & \frac{p\left(A_{i}\wedge a_{i}\wedge\lambda\right)}{p\left(a_{i}\wedge\lambda\right)}\,\frac{p\left(B_{j}\wedge b_{j}\wedge\lambda\right)}{p\left(b_{j}\wedge\lambda\right)}\end{eqnarray*}
So, parameter $\lambda$, standing for the Cauchy data carrying the
information shared by the left and right wings, must satisfy the following
so-called {}``screening off'' condition: \begin{equation}
p\left(A_{i}\wedge B_{j}|a_{i}\wedge b_{j}\wedge\lambda\right)=p\left(A_{i}|a_{i}\wedge\lambda\right)p\left(B_{j}|b_{j}\wedge\lambda\right)\label{eq_Bell_szorzat}\end{equation}

Bell restricted the concept of LDM embedding with a further requirement
which is nothing but Assumption~4. In this context it says the following:
The choice between the possible measurement setups must be independent
from parameter $\lambda$ carrying the shared information. In other
words,\begin{equation}
\begin{array}{ccc}
u^{a_{i}}\left(\mu,\lambda\right) & = & u^{a_{i}}\left(\mu\right)\\
u^{b_{i}}\left(\lambda,\nu\right) & = & u^{b_{i}}\left(\nu\right)\end{array}\,\,\,\, i=1,2\label{eq_no_conspiracy}\end{equation}

In this case, it immediately follows from (\ref{eq_szallas1})--(\ref{eq_szallas6})
that \begin{eqnarray}
p(A_{i}|a_{i}) & = & \sum_{\lambda}p\left(A_{i}|a_{i}\wedge\lambda\right)p\left(\lambda\right)\nonumber \\
p(B_{i}|b_{i}) & = & \sum_{\lambda}p\left(B_{i}|b_{i}\wedge\lambda\right)p\left(\lambda\right)\,\,\,\,\,\,\,\,\,\,\,\,\,\,\,\, i,j=1,2\label{eq_Bell_teljes}\\
p(A_{i}\wedge B_{j}|a_{i}\wedge b_{j}) & = & \sum_{\lambda}p\left(A_{i}\wedge B_{j}|a_{i}\wedge b_{j}\wedge\lambda\right)p\left(\lambda\right)\nonumber \end{eqnarray}
For example:\begin{eqnarray*}
p\left(A_{i}|a_{i}\right) & = & \frac{p\left(A_{i}\wedge a_{i}\right)}{p\left(a_{i}\right)}=\frac{\sum_{\mu,\lambda}u^{A_{i}}\left(\mu,\lambda\right)u^{a_{i}}\left(\mu,\lambda\right)p\left(\mu\right)p\left(\lambda\right)}{\sum_{\mu,\lambda}u^{a_{i}}\left(\mu,\lambda\right)p\left(\mu\right)p\left(\lambda\right)}\\
 & = & \frac{\sum_{\mu,\lambda}u^{A_{i}}\left(\mu,\lambda\right)p\left(\mu\right)p\left(\lambda\right)}{\sum_{\mu,\lambda}u^{a_{i}}\left(\mu,\lambda\right)p\left(\mu\right)p\left(\lambda\right)}=\frac{\sum_{\lambda}\left(\sum_{\mu}u^{A_{i}}\left(\mu,\lambda\right)p\left(\mu\right)\right)p\left(\lambda\right)}{\sum_{\mu}u^{a_{i}}\left(\mu\right)p\left(\mu\right)}\\
 & \begin{array}[b]{c}
(\star)\\
=\end{array} & \sum_{\lambda}\left(\frac{\sum_{\mu}u^{A_{i}}\left(\mu,\lambda\right)p\left(\mu\right)}{\sum_{\mu}u^{a_{i}}\left(\mu\right)p\left(\mu\right)}\right)p\left(\lambda\right)=\sum_{\lambda}p\left(A_{i}|a_{i}\wedge\lambda\right)p\left(\lambda\right)\end{eqnarray*}
Equality $(\star)$ would not hold without condition (\ref{eq_no_conspiracy}).

It is an elementary fact that for any real numbers $0\leq x_{1},x_{2},y_{1},y_{2}\leq1$\[
-1\leq x_{1}y_{1}+x_{1}y_{2}+x_{2}y_{2}-x_{2}y_{1}-x_{1}-y_{2}\leq0\]
 Applying this inequality, for all $\lambda$ we have\[
\begin{array}{r}
-1\leq p\left(A_{1}|a_{1}\wedge\lambda\right)p\left(B_{1}|b_{1}\wedge\lambda\right)+p\left(A_{1}|a_{1}\wedge\lambda\right)p\left(B_{2}|b_{2}\wedge\lambda\right)\\
+p\left(A_{2}|a_{2}\wedge\lambda\right)p\left(B_{2}|b_{2}\wedge\lambda\right)-p\left(A_{2}|a_{2}\wedge\lambda\right)p\left(B_{1}|b_{1}\wedge\lambda\right)\\
-p\left(A_{1}|a_{1}\wedge\lambda\right)-p\left(B_{2}|b_{2}\wedge\lambda\right)\leq0\end{array}\]
Taking into account (\ref{eq_Bell_szorzat}), we obtain: \begin{equation}
\begin{array}{r}
-1\leq p\left(A_{1}\wedge B_{1}|a_{1}\wedge b_{1}\wedge\lambda\right)+p\left(A_{1}\wedge B_{2}|a_{1}\wedge b_{2}\wedge\lambda\right)\\
+p\left(A_{2}\wedge B_{2}|a_{2}\wedge b_{2}\wedge\lambda\right)-p\left(A_{2}\wedge B_{1}|a_{2}\wedge b_{1}\wedge\lambda\right)\\
-p\left(A_{1}|a_{1}\wedge\lambda\right)-p\left(B_{2}|b_{2}\wedge\lambda\right)\leq0\end{array}\label{eq_Bell_CH_levezetes1}\end{equation}
Multiplying this with probability $p\left(\lambda\right)$ and summing
up over $\lambda$, we obtain the following inequality:

\begin{equation}
\begin{array}{r}
-1\leq p\left(A_{1}\wedge B_{1}|a_{1}\wedge b_{1}\right)+p\left(A_{1}\wedge B_{2}|a_{1}\wedge b_{2}\right)\\
+p\left(A_{2}\wedge B_{2}|a_{2}\wedge b_{2}\right)-p\left(A_{2}\wedge B_{1}|a_{2}\wedge b_{1}\right)\\
-p\left(A_{1}|a_{1}\right)-p\left(B_{2}|b_{2}\right)\leq0\end{array}\label{eq_Bell_CH1}\end{equation}
Similarly, changing the roles of $A_{1},A_{2},B_{1}$, and $B_{2}$,
we have:

\begin{equation}
\begin{array}{r}
-1\leq p\left(A_{2}\wedge B_{1}|a_{2}\wedge b_{1}\right)+p\left(A_{2}\wedge B_{2}|a_{2}\wedge b_{2}\right)\\
+p\left(A_{1}\wedge B_{2}|a_{1}\wedge b_{2}\right)-p\left(A_{1}\wedge B_{1}|a_{1}\wedge b_{1}\right)\\
-p\left(A_{2}|a_{2}\right)-p\left(B_{2}|b_{2}\right)\leq0\end{array}\label{eq_Bell_CH2}\end{equation}
\begin{equation}
\begin{array}{r}
-1\leq p\left(A_{1}\wedge B_{2}|a_{1}\wedge b_{2}\right)+p\left(A_{1}\wedge B_{1}|a_{1}\wedge b_{1}\right)\\
+p\left(A_{2}\wedge B_{1}|a_{2}\wedge b_{1}\right)-p\left(A_{2}\wedge B_{2}|a_{2}\wedge b_{2}\right)\\
-p\left(A_{1}|a_{1}\right)-p\left(B_{1}|b_{1}\right)\leq0\end{array}\label{eq_Bell_CH3}\end{equation}
\begin{equation}
\begin{array}{r}
-1\leq p\left(A_{2}\wedge B_{2}|a_{2}\wedge b_{2}\right)+p\left(A_{2}\wedge B_{1}|a_{2}\wedge b_{1}\right)\\
+p\left(A_{1}\wedge B_{1}|a_{1}\wedge b_{1}\right)-p\left(A_{1}\wedge B_{2}|a_{1}\wedge b_{2}\right)\\
-p\left(A_{2}|a_{2}\right)-p\left(B_{1}|b_{1}\right)\leq0\end{array}\label{eq_Bell_CH4}\end{equation}

\noindent Inequalities (\ref{eq_Bell_CH1})--(\ref{eq_Bell_CH4})
are due to Clauser and Horne \cite{key-39}, but they essentially
play the same role as Bell's original inequalities of 1964. Therefore
they are called \emph{Bell--Clauser--Horne} inequalities. 

According to Assumption~1, the conditional probabilities in the Bell--Clauser--Horne
inequalities are nothing but the corresponding quantum probabilities,
the values of which are given in (\ref{eq_EPR_szamok1})--(\ref{eq_EPR_szamok3}).
These values violate the Bell--Clauser--Horne inequalities.

\emph{So, in a different context, we arrived at conclusions similar
to section~\ref{sub:The-EPR-conclusion}. That is to say, one of
Assumption~1, Assumption~2' and Assumption~4 must fail.}

Notice that the Clauser--Horne--Pitowsky inequalities (\ref{eq_CH-Pitowsky})
and the Bell--Clauser--Horne inequalities (\ref{eq_Bell_CH1})--(\ref{eq_Bell_CH4})
are not identical---in spite of the obvious similarity. The formers
apply to some numbers that are meant to be the \emph{(absolute) probabilities}
of particular events, and express the necessary condition of that
these {}``probabilities'' admit a Kolmogorovian representation and---in
the Laboratory Record Argument---a relative frequency interpretation.
In contrast the Bell--Clauser--Horne inequalities apply to \emph{conditional
probabilities}, and we derived them as necessary conditions of LDM
embedability. 

Finally, it worthwhile mentioning, that the spin-correlation experiment
described in section~\ref{sub:The-description-of}  has been performed
in reality, partly with spin-$\frac{1}{2}$ particles, partly with
photons \cite{key-33}. (The experimental scenario for spin-$\frac{1}{2}$
particles can easily be translated into the terms of polarization
measurements with entangled photon pairs.) In the experiments with
photons, the spatial separation of the left and right wing measurements
has also been realized. (The first experiment in which the spatial
separation was realized is \cite{key-34}. The best conditions have
been achieved in \cite{key-35}.) So far, the experimental results
have been in wonderful agreement with quantum mechanical predictions.
Therefore, \emph{the violation of the Bell-type inequalities is an
experimental fact.}

In the particular case when the values of $p\left(A_{i}|a_{i}\wedge\lambda\right)$,
$p\left(B_{i}|b_{i}\wedge\lambda\right)$, and $p\left(A_{i}\wedge B_{j}|a_{i}\wedge b_{j}\wedge\lambda\right)$
on the right hand side of (\ref{eq_Bell_teljes}) are only $0$ or
$1$, $\lambda$ is called a \emph{deterministic hidden variable}.
The above derivation of the Bell--Clauser--Horne inequalities simultaneously
holds for both stochastic and deterministic hidden variable theories.
Notice that the screening off condition (\ref{eq_Bell_szorzat}) is
not automatically satisfied by any deterministic hidden variable.
What we automatically have in the deterministic case is the following:
\[
p\left(A_{i}\wedge B_{j}|a_{i}\wedge b_{j}\wedge\lambda\right)=p\left(A_{i}|a_{i}\wedge b_{j}\wedge\lambda\right)p\left(B_{j}|a_{i}\wedge b_{j}\wedge\lambda\right)\]
This is different from condition (\ref{eq_Bell_szorzat}), except
if the following are also satisfied: \begin{eqnarray}
p\left(A_{i}|a_{i}\wedge b_{j}\wedge\lambda\right) & = & p\left(A_{i}|a_{i}\wedge\lambda\right)\label{eq_parameter_fuggetlenseg1}\\
p\left(B_{j}|a_{i}\wedge b_{j}\wedge\lambda\right) & = & p\left(B_{j}|b_{j}\wedge\lambda\right)\label{eq_parameter_fuggetlenseg2}\end{eqnarray}
that is to say, the outcome in the left wing is independent of the
choice of the measurement setup in the right wing, and \emph{vice
versa}. Conditions (\ref{eq_parameter_fuggetlenseg1})--(\ref{eq_parameter_fuggetlenseg2}),
sometimes called {}``parameter independence'' \cite{key-54}, are,
however, automatically satisfied by LDM embedability. 

Thus, the distinction between deterministic and stochastic hidden
variable theories is not so significant. As we have seen, the necessary
condition of their existence is common to both of them.

When we say that the hidden variable model is {}``stochastic'',
it means \emph{epistemic} stochasticity. Parameter $\lambda$ does
not fully determine the measurement outcomes: the value of $u^{A_{i}}\left(\mu,\lambda\right)$
also depends on $\mu$, and the value of $u^{B_{j}}\left(\lambda,\nu\right)$
also depends on $\nu$. But the LDM world, as a whole, is deterministic:
whether events $A_{i}$ and $B_{j}$ occur is fully determined by
$\mu$, $\lambda$, and $\nu$.

\section{Possible resolutions of the paradox}

\subsection{Conspiracy}

There is an easy resolution of the EPR/Bell paradox, if we allow the
conspiracy that was prohibited by Assumption~4 \cite{key-32,key-5}.
It is hard to believe, however, that the {}``free'' decisions of
the laboratory assistants in the left and right wings depend on the
value of the hidden variable which also determines the spins of the
two particles.

\subsection{Fine's interpretation of quantum statistics}

\begin{figure}
\begin{centering}
\includegraphics[width=0.7\columnwidth]{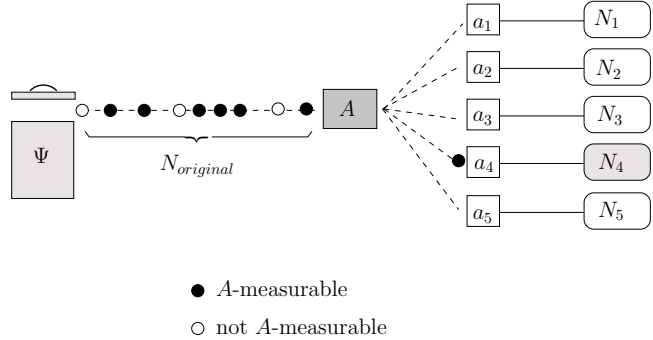}
\par\end{centering}

\caption{\emph{The schema of a typical quantum measurement. The source is producing
objects on which the measurement is performed. The very existence
of an object can be observed via the detection of an outcome event.
Therefore, we have no information about the content of the original
ensemble of objects emitted by the source. The quantum probabilities
are identified with the frequencies of the different outcomes, relative
to a sub-ensemble of objects producing any outcome\label{fig_qmeasurement}}}

\end{figure}

Assumption~1 seems to be the most robust one. One might think that
(\ref{eq_mi_trace_min}) is a simple empirical fact. There is, however,
a resolution of the problem which is entirely compatible with Assumptions~2'
and 4, but violates Assumption~1 in a very a sophisticated way. This
is Arthur Fine's interpretation of quantum statistics \cite{key-13}.
The basic idea is this. To determine `What does quantum probability
actually describe in the real world?' we have to analyze the actual
empirical counterpart of $tr\left(\hat{W}P_{i}\right)$ in the experimental
confirmations of quantum theory. Consider the schema of a typical
quantum measurement (Fig.~\ref{fig_qmeasurement}). %
 Contrary to classical physics where getting information about the
existence of a physical entity and measuring one of its characteristics
are two different actions, in a typical quantum measurement these
two actions coincide. Therefore we have no independent information
about the content of the original ensemble of objects emitted by the
source. In fact, the theoretical {}``probability'' predicted by quantum
mechanics is identified with the ratio of the number of detections
in one channel relative to the total number of detections, that is,

\begin{equation}
tr\left(\hat{W}P_{i}\right)=\frac{N_{i}}{\sum_{i}N_{i}}\label{eq:arany}\end{equation}

Now, if, as it is usually assumed, a non-detection were an \emph{independent}
random mistake of an inefficient detector or something like that,
then the right hand side of (\ref{eq:arany}) would be still equal
to $p\left(<a_{i}>|a\right)$. This is, however, a completely implausible
assumption within the context of a hidden variable theory. (This is
the most essential point of Fine's approach.) For if there are (hidden)
elements of reality, for instance the particle has some hidden properties,
that pre-determine the outcome of the measurement and in general pre-determine
the behavior of the system during the whole measurement process, then
it is quite plausible that they also pre-determine whether the entity
in question can pass through the analyzer and can be detected, or
not. If so, then the right hand side of (\ref{eq:arany}) is a relative
frequency on a {}``biased'' ensemble, therefore \[
p\left(\widetilde{<a_{i}>}\right)=p\left(<a_{i}>|a\right)\neq tr\left(\hat{W}P_{i}\right)\]

\noindent and the the Clauser--Horne--Pitowsky inequalities as well
as the Bell--Clauser--Horne inequalities can be---and, in fact, are---satisfied.
This is, of course, not the whole story. The concrete hidden variable
theory has to describe how the hidden properties determine the whole
process and how the relative frequencies of the hidden elements of
reality are related to quantum probabilities. \emph{There exist such
hidden variable models for several spin-correlation experiments and
they are entirely compatible with the real experiments performed in
the last few years.} For further reading see \cite{key-8,key-14,key-17,key-22,key-19}.

\subsection{Non-locality, but without communication}

In spite of the above mentioned developments and in spite of the fact
that the no-action-at-a-distance principle seems to hold in all other
branches of physics, the painful conclusion that Assumption~2 is
violated is more widely accepted in contemporary philosophy of physics.

Many argue that the violation of locality observed in the EPR experiment
is not a serious one, because the spin-correlations are not capable
of transmitting information between spatially separated space-time
regions. The argument is based on the fact that, although the outcome
in the right wing is (maximally) correlated with the outcome in the
left wing, the outcome in the left wing itself is a random event (with
probability $\frac{1}{2}$ it is `up' or `down') which cannot be influenced
by our free action. We cannot send Morse code signals from the left
station to the right one with an EPR equipment.

\begin{figure}[h]
\noindent %
\begin{minipage}[c][1\totalheight]{0.1\textwidth}%
\begin{center}
(A)
\par\end{center}%
\end{minipage}%
\begin{minipage}[c][1\totalheight]{0.9\textwidth}%
\begin{center}
\includegraphics[width=0.8\columnwidth]{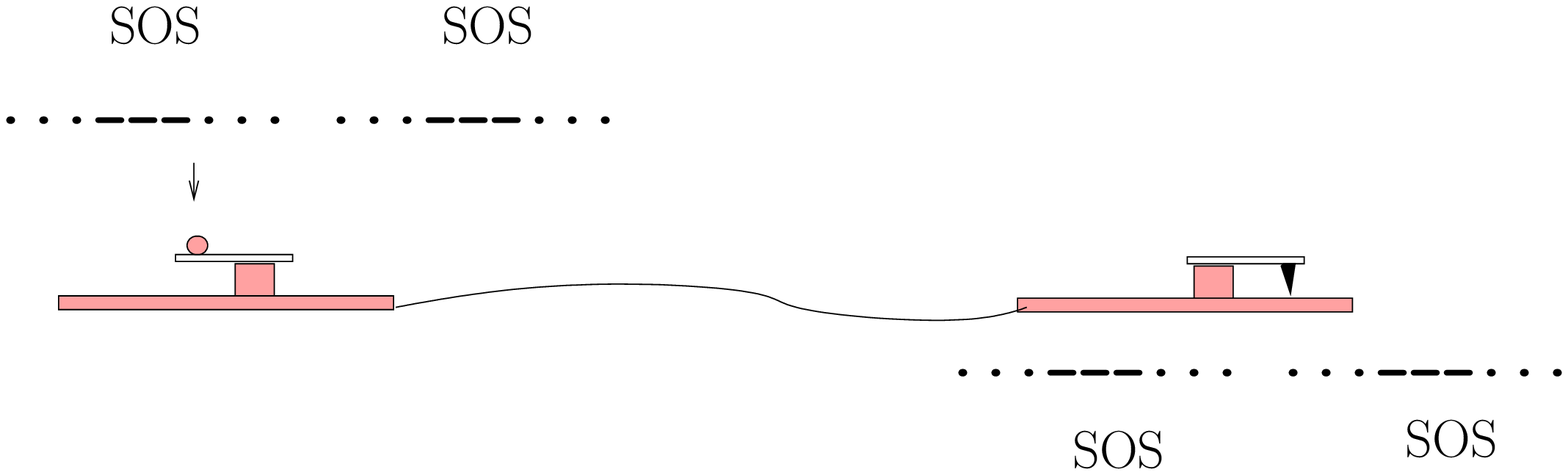}
\par\end{center}%
\end{minipage}%

\noindent %
\begin{minipage}[c][1\totalheight]{0.1\textwidth}%
\begin{center}
(B)
\par\end{center}%
\end{minipage}%
\begin{minipage}[c][1\totalheight]{0.9\textwidth}%
\begin{center}
\includegraphics[width=0.8\columnwidth]{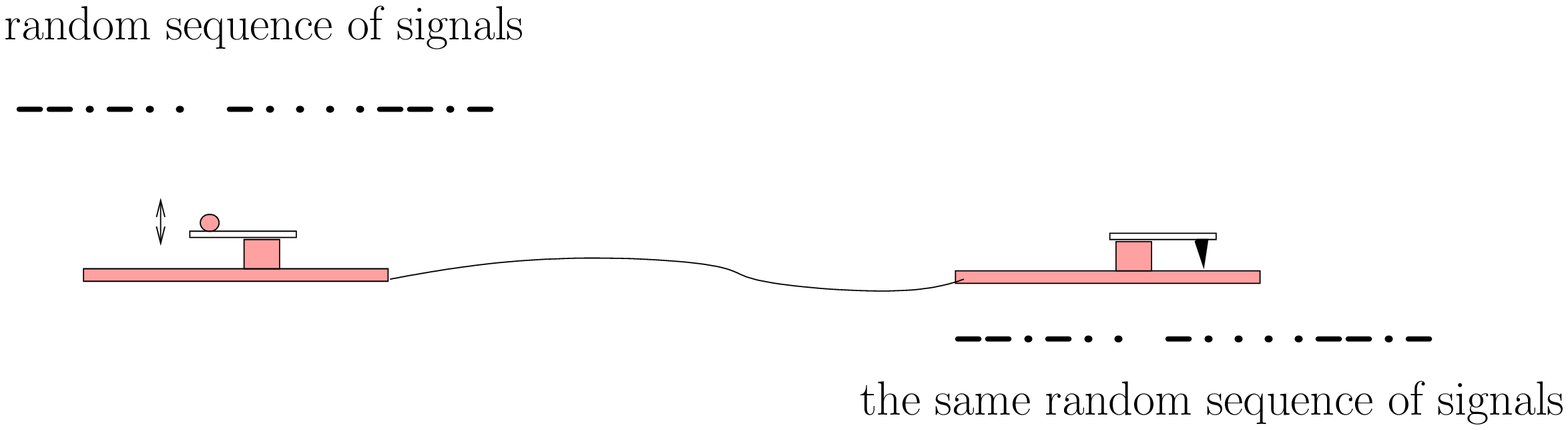}
\par\end{center}%
\end{minipage}%

\noindent %
\begin{minipage}[c][1\totalheight]{0.1\textwidth}%
\begin{center}
(C)
\par\end{center}%
\end{minipage}%
\begin{minipage}[c][1\totalheight]{0.9\textwidth}%
\begin{center}
\includegraphics[width=0.8\columnwidth]{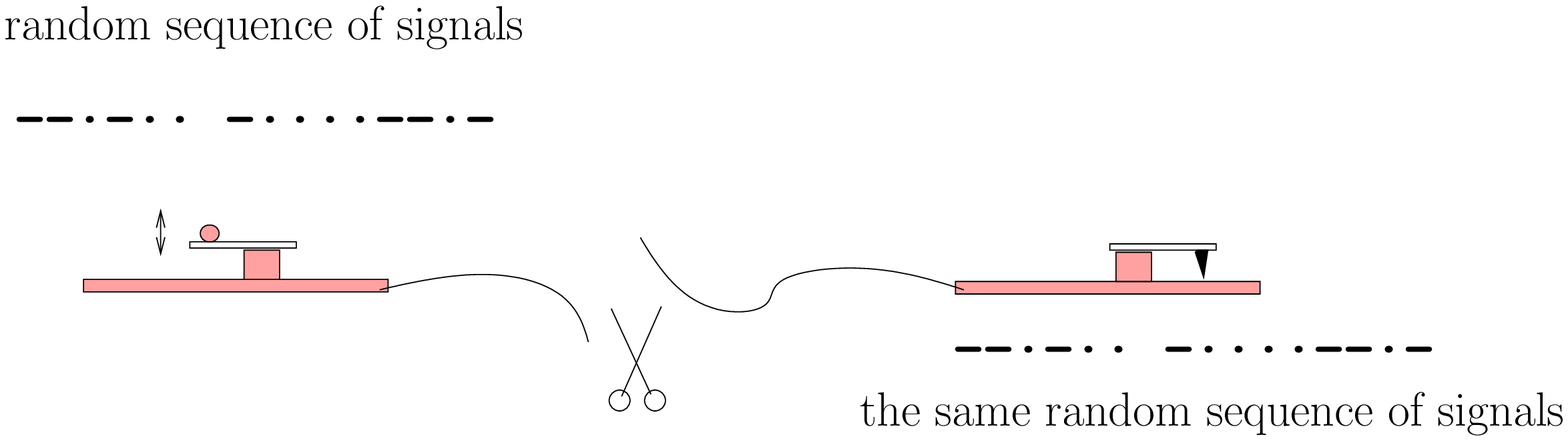}
\par\end{center}%
\end{minipage}%

\caption{\emph{In case (A) the telegraph works normally. In case (B) something
goes wrong and the key randomly presses itself. The random signal
is properly transmitted but the equipment is not suitable for sending
a telegram. Case (C) is just like (B), but the cable connecting the
two equipments is broken\label{fig_taviro}}}

\end{figure}

Others argue that this is a misinterpretation of the original no-action-at-a-distance
principle which completely prohibits spatially separated physical
events having any causal influence on each other, no matter whether
or not the whole process is suitable for transmission of information.
Consider the example depicted in Fig.~\ref{fig_taviro}. %
 In case (A) the telegraph works normally. By pressing the key we
can send information from one station to the other. It is no wonder
that the pressing of the key at the sender station and the behavior
of the register at the receiver station are maximally correlated.
We have a clear causal explanation of how the signal is propagating
along the cable connecting the two stations. Next, imagine that something
goes wrong and the key randomly presses itself (case (B)). The random
sequence of signals generated in this way is properly transmitted
to the receiver station, but the system is not suitable to send telegrams.
Still we have a clear causal explanation of the correlation between
the behaviors of the key and the register. Finally, case (C), imagine
the same situation as (B) except that the cable connecting the two
stations is broken. In this situation, it would be astonishing if
there really were correlations between the random behavior of the
key and the behavior of the register, and it would cry out for causal
explanation, no matter whether or not we are able to send information
from one station to the other.

As this simple example illustrates, no matter whether or not we are
able to communicate with EPR equipment, the very fact that we observe
correlations which cannot be accommodated in the causal order of the
world is still an embarrassing metaphysical problem.

\subsection{Modifying the theory}

In order to resolve the paradox, there have been various suggestions
to modify the underlying physical/mathematical/logical theories by
which we describe the phenomena in question. Some of these endeavors
are based on the observation that the violation of the Bell-type inequalities
is deeply related to the non-classical feature of quantum probability
theory \cite{key-40,key-4,key-41,key-42}. More exactly, it is rooted
in the (non-distributive lattice) structure of the underlying event
algebra which essentially differs from the classical Boolean algebra.
According to some of these approaches, the fact in itself that the
Bell-type inequalities are violated has nothing to do with such physical
questions as locality, causality or the ontology of quantum phenomena.
It is just a simple mathematical consequence of quantum probability
theory and/or quantum logic (\cite{key-4}, pp. 49--51; 182--183). 

According to another approach, it is quantum mechanics itself that
has to be modified. So called relational quantum mechanics \cite{key-43,key-44,key-45}
introduces a new concept: the relative quantum state. It turns out
that the relative quantum state of the right particle changes if the
left particle is measured and \emph{vice versa}. Therefore, it is
argued, the two particles are not causally separated at a quantum
level.

Some papers, motivated by the problem of quantum gravity, suggest
space-time structures that are intrinsically based on quantum theory.
These results have remarkable interrelations with the EPR--Bell problem
\cite{key-46,key-47,key-48}. The EPR events, which are spatially
separated in classical space-time, turn out not to be spatially separated
in some other space-time structures based on quantum mechanics.

Another branch of research attempts to develop, within the framework
of algebraic quantum field theory, an exact concept of {}``separation''
of subsystems \cite{key-52,key-49,key-51,key-50}.

What is common to all these efforts is that they aim to improve the
conceptual/theoretical means by which we describe and analyze the
EPR--Bell problem. All these approaches, however, encounter the following
difficulty: The violation of the Bell-type inequalities is an \emph{experimental
fact}. It means that the EPR--Bell problem exists independently of
quantum mechanics, and independently of any other \emph{theories}:
what is important from (\ref{eq_EPR_prob_fel})--(\ref{eq_EPR_prob_sin})
is that

\begin{eqnarray}
p(A|a)=p(B|b) & = & \frac{1}{2}\label{eq_EPR_prob_fel-lenyeg}\\
p(A\wedge B|a\wedge b) & = & \frac{1}{2}\sin^{2}\frac{\sphericalangle(\mathbf{a,b)}}{2}\label{eq_EPR_prob_sin-lenyeg}\end{eqnarray}
We \emph{observe} correlations in the macroscopic world, which have
no satisfactory explanation. It is hard to see how we could resolve
the EPR--Bell paradox by changing something in our theories, by introducing
new concepts, by changing, for example, the notion of a quantum state,
by applying {}``quantum logic'', {}``quantum space-time'', etc.
For, \emph{until the modified theory can reproduce the experimentally
observed relative frequencies (\ref{eq_EPR_prob_fel-lenyeg})--(\ref{eq_EPR_prob_sin-lenyeg}),
the modified theory will contradict to Assumptions 1, 2/2', and 4.}
(Note that Fine's approach differs from the other proposals in claiming
that (\ref{eq_EPR_prob_fel-lenyeg})--(\ref{eq_EPR_prob_sin-lenyeg})
are not what we actually observe in the real experiments).

\section{No correlation without causal explanation\label{sec:No-correlation-without}}

How correlations between event types are related to causality between
particular events is an old problem in the history of philosophy.
Although the underlying causality on the level of particular events
does not necessarily yield to correlations on the level of event types,
it is a deeply rooted metaphysical conviction, on the other hand,
that \emph{there is no correlation without causal explanation}. If
there is correlation between two event types then there must exist
something in the common causal past of the corresponding particular
events that explains the correlation. This something is called a {}``common
cause''. {}``Particular event'' means an event of a definite space-time
locus, a definite piece of the history of the universe, that is the
totally detailed state of affairs in a given space-time region. 

The interesting situation is, of course, when the correlated events
are not in direct causal relationship; for example, they are simultaneous
or, at least, spatially separated. (In order to distinguish direct
causal relations from common-cause-type causal schemas, in other words
real causal processes from pseudo-processes, Reichenbach \cite{key-37}
and Salmon \cite{key-38} introduced the so called mark-transmission
criterion: a direct causal process is capable of transmitting a local
modification in structure (a {}``mark''); a pseudo-process is not.
Consider Salmon's simple example: as the spotlight rotates, the spot
of light moves around the wall. We can place a red filter at the wall
with the result that the spot of light becomes red at that point.
But if we make such a modification in the travelling spot, it will
not be transmitted beyond the point of interaction. The {}``motion''
of the spot of light on the wall is not a real causal process. On
the contrary, the propagation of light from the spotlight to the wall
is a real causal process. If we place a red filter in front of the
spotlight, the change of color propagates with the light signal to
the wall, and the spot of light on the wall becomes red. It is not
entirely clear, however, how the mark-transmission criterion is applicable
for objectively random uncontrollable phenomena, like the EPR experiment.
It also must be mentioned that the criterion is based on some prior
metaphysical assumptions about free will and free action.)

The idea that a correlation between events having no direct causal
relation must always have a common-cause explanation is due to Hans
Reichenbach \cite{key-37}. It is hotly disputed whether the principle
holds at all. Many philosophers claim that there are {}``regularities''
in our world that have no causal explanations. The most famous such
example was given by Elliot Sober \cite{key-53}: The bread prices
in Britain have been going up steadily over the last few centuries.
The water levels in Venice have been going up steadily over the last
few centuries. There is therefore a {}``regularity'' between simultaneous
bread prices in Britain and sea levels in Venice. However, there is
presumably no direct causation involved, nor a common cause. Of course,
{}``regularity'' here does not mean \emph{correlation} in probability-theoretic
sense ($p(A\wedge B)-p(A)p(B)=1\cdot1-1=0$). So, it is still an open
question whether the principle holds, in its original Reichenbachian
sense, for events having non-zero correlation. Various examples from
classical physics have been suggested which violate Reichenbach's
common cause principle. There is no consensus on whether these examples
are valid. There is, however, a consensus that the EPR--Bell problem
is a serious challenge to Reichenbach's principle.

Another much-discussed problem is how to define the concept of common
cause. As we have seen, in Bell's understanding, the common cause
is the hidden state of the universe in the intersection of the backward
light cones of the correlated events. This view is based on the LDM
world view of the pre-quantum-mechanical physics. According to Reichenbach's
definition (\cite{key-37}, Chapter~19) a common cause explaining
the correlation $p(A\wedge B)-p(A)p(B)\neq0$ is an event $C$ satisfying
the following condition:\begin{eqnarray}
p\left(A\wedge B|C\right) & = & p\left(A|C\right)p\left(B|C\right)\label{eq:Reichenbach1}\\
p\left(A\wedge B|\neg C\right) & = & p\left(A|\neg C\right)p\left(B|\neg C\right)\label{eq:Reichenbach2}\end{eqnarray}
Reichenbach based his common-cause concept on intuitive examples from
the classical world with epistemic probabilities. However, as Nancy
Cartwright \cite{key-36} points out, we are in trouble if the world
is objectively indeterministic. We have no suitable metaphysical language
to tell when a world is local, to tell the difference between direct
and common-cause-type correlations, to tell what a common cause is,
and so on. These concepts of the theory of stochastic causality are
either unjustified or originated from the observations of epistemically
stochastic phenomena of a deterministic world.

\end{document}